\documentclass{aa}  

\usepackage{graphicx}
\usepackage{txfonts}
\usepackage[colorlinks=false]{hyperref}
\newcommand{\RNum}[1]{\uppercase\expandafter{\romannumeral #1\relax}}
\newcommand{\Fermitool}{{\fontfamily{qcr}\selectfont FermiTool}}
\newcommand{\IRF}{{\fontfamily{qcr}\selectfont P8R3\_SOURCE\_V3}}
\newcommand{\isotropic}{{\fontfamily{qcr}\selectfont iso\_P8R3\_SOURCE\_V3\_v1}}
\newcommand{\galactic}{{\fontfamily{qcr}\selectfont gll\_iem\_v07}}
\newcommand{\srcprob}{{\fontfamily{qcr}\selectfont gtsrcprob}}

\begin{document}

   \title{Detection of minute-timescale $\gamma-$ray variability in BL Lacertae by {\it Fermi}-LAT}


   \author{A. Pandey\thanks{ashwanitapan@gmail.com}
          \and
          C. S. Stalin
          }

   \institute{Indian Institute of Astrophysics, Block II, Koramangala, Bangalore 560034, India}


 
  \abstract
 {BL Lacertae, the prototype of the BL Lacertae (BL Lac) category of blazars, underwent a giant $\gamma-$ray flare in April 2021. The Large Area Telescope (LAT) onboard the {\it Fermi} Gamma-ray Space Telescope (hereafter {\it Fermi}-LAT) observed a peak $\gamma-$ray (0.1$-$500 GeV) flux of $\sim$2 $\times$ 10$^{-5}$ photons cm$^{-2}$ s$^{-1}$ within a single orbit on 2021 April 27, which is historically the brightest $\gamma-$ray flux ever detected from the source. Here, we report, for the first time, the detection of significant minute-timescale GeV  $\gamma-$ray flux variability in the BL Lac subclass of blazars by the {\it Fermi}-LAT. We resolved the source variability down to 2-min binned timescales with a flux halving time of $\sim$1 minute, which is the shortest GeV variability timescale ever observed from blazars.  The detected variability timescale is much shorter than the light-crossing time ($\sim  14$ minutes) across the central black hole of BL Lac indicating a very compact $\gamma-$ray emission site within the outflowing jet. Such a compact emitting region requires the bulk Lorentz factor of the jet to be larger than 16 so that the jet power is not super Eddington. We found a minimum Doppler factor $\delta_{min}$ of 15 using the $\delta$ function approximation for the $\gamma\gamma$ opacity constraint. For a conical jet geometry, considering $\Gamma = \delta_{min}$, the observed short variability timescale suggests the very compact emission region to lie at a distance of about 8.62 $\times$ 10$^{14}$ cm from the central engine of BL Lac.}


   \maketitle
%

\section{Introduction}
The extragalactic $\gamma$-ray sky is dominated by the blazar category of active galactic nuclei (AGN; \citealt{2020ApJS..247...33A}). Blazars comprising flat-spectrum radio quasars (FSRQs) and BL Lac objects (BL Lacs) are radio-loud AGN that have their relativistic jets aligned close to the line of sight to the observer \citep{1978bllo.conf..328B,1995PASP..107..803U}. They emit most of their energy in high energy $\gamma$-rays with the most powerful sources, during strong flares reaching  $\gamma$-ray luminosities (in the isotropic emission scenario) as high as $L_{\gamma} \sim 10^{49-50}$ erg s$^{-1}$ erg s$^{-1}$ \citep{2017ApJ...841...61A}. The broadband spectral energy distributions (SEDs) of blazars have  double-hump structures \citep{1998MNRAS.299..433F}. The low-energy component of the SED is well explained by synchrotron emission from the relativistic electrons within the jet, while the origin of the high-energy component is still unclear \citep[e.g.][]{2007Ap&SS.307...69B}. One among the mechanisms responsible for the high energy $\gamma$-ray radiation is the synchrotron self Compton (SSC) process by which the synchrotron photons emitted by the relativistic electrons in the jet are Compton up-scattered by the same population of electrons in the jet \citep{1992ApJ...397L...5M}. The other model posits the production of $\gamma$-rays by inverse Compton scattering of seed photons external 
to the jet, by electrons in the emitting jet. The seed photons could be the ultra-violet photons from the accretion disk \citep{1993ApJ...416..458D}, the photons from emission lines from the broad-line region \citep{1994ApJ...421..153S} and the infrared emission from the dusty torus \citep{2000ApJ...545..107B}. Alternatively, hadronic processes could also produce $\gamma$-rays \citep{2013ApJ...768...54B}.
Based on the synchrotron peak frequency ($\nu_{peak}$) of their SEDs, BL Lacs are further divided into low-frequency peaked BL Lacs (LBLs; $\nu_{peak} \leq 10^{14}$ Hz), intermediate-frequency peaked BL Lacs (IBLs; $10^{14}$ Hz $< \nu_{peak} < 10^{15}$Hz), and high-frequency peaked BL Lacs (HBLs; $\nu_{peak} \geq 10^{15}$Hz) \citep{1995ApJ...444..567P,2010ApJ...716...30A}.

The $\gamma-$rays from blazars are known to be highly variable \citep{2010ApJ...722..520A,2020A&A...634A..80R} which indicates the emission region is highly compact \citep{1994ApJS...94..551F} and the $\gamma$-ray radiation is highly beamed \citep{1995MNRAS.273..583D}. However, the exact physical processes that are responsible for the generation of the observed $\gamma$-ray emission, as well as their production site, are uncertain and highly debated. Using powerful observational facilities in the $\gamma-$ray domain, rapid $\gamma-$ray variations have been detected in 
seven AGN so far, including three BL Lacs (PKS 2155-304 \citep{2007ApJ...664L..71A}, Mrk 501 \citep{2007ApJ...669..862A}, BL Lac \citep{2013ApJ...762...92A,2019A&A...623A.175M}), three FSRQs (PKS PKS 1222$+$21 \citep{2011ApJ...730L...8A}, 3C 279 \citep{2016ApJ...824L..20A}, CTA 102 \citep{2016ApJ...824L..20A}), and one radio galaxy (IC 310 \citep{2014Sci...346.1080A}. However, the minute-timescale $\gamma-$ray variations were also detected multiple times in the same source. For example, \cite{2013ApJ...762...92A} observed a rapid TeV $\gamma-$ray flare from BL Lac with an exponential decay time of 13$\pm$4 min and recently, \cite{2019A&A...623A.175M} also reported the detection of VHE $\gamma-$ray flare  with halving time of 26$\pm$8 min from the same source. Moreover, the majority of the findings of rapid $\gamma$-ray variations come from VHE observations by the ground-based Cerenkov telescopes except for two sources, namely 3C 279 \citep{2016ApJ...824L..20A} and CTA 102 \citep{2018ApJ...854L..26S}, where the short time scale variations are from the GeV observations by the {\it Fermi}-LAT. Both of these sources belong to the FSRQ subclass of blazars. To date, no rapid GeV $\gamma-$ray variability was detected in any BL Lacs by the {\it Fermi}-LAT.
\begin{figure*}
\centering
\includegraphics[width=15.5cm, height=13cm]{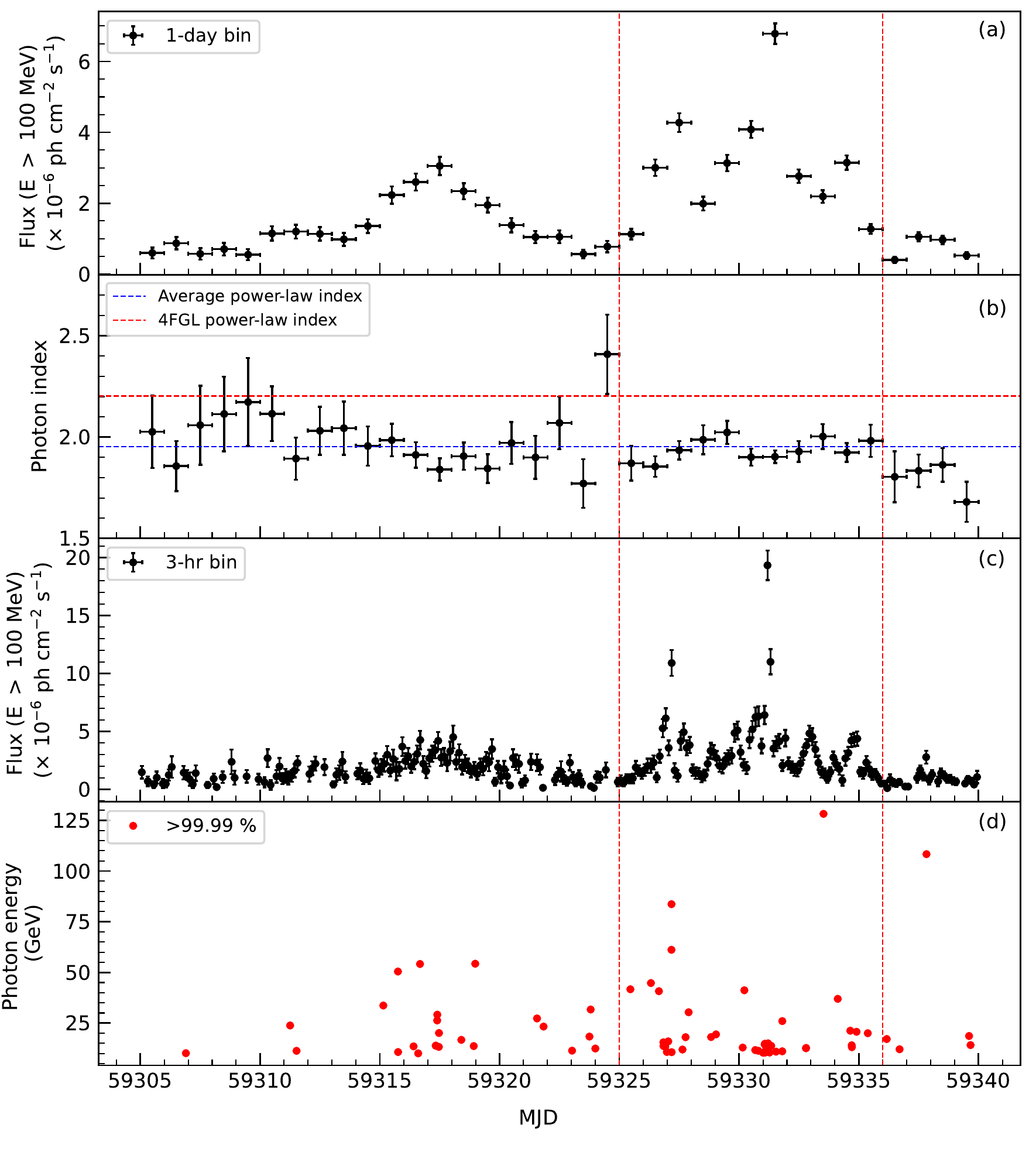}
\caption{\label{fig:lc_lat} Panel (a): The one-day binned light curve of BL Lac in the $\gamma-$ray band (0.1$-$500 GeV) covering a period of $\sim$ 35 days. Panel (b): The variation of $\gamma-$ray photon index with time; the blue and red horizontal lines indicate the average power-law index during the period considered and the power-law index mentioned in the 4FGL catalog, respectively. Panel (c): The 3-hr binned $\gamma-$ray light curve of BL Lac. Panel (d): The arrival time distribution of photons with energies greater than 10 GeV. The vertical red lines indicate the period of the major $\gamma-$ray flare.}
\end{figure*}
The availability of data from the {\it Fermi}-LAT \citep{2009ApJ...697.1071A} observations provide ample opportunities to find evidence of short time scale of variations in more blazars. Our motivation here is to find rapid $\gamma-$ray variations in the BL Lac subclass of blazars using the {\it Fermi}-LAT observations. BL Lac, at a redshift of $z$=0.069 \citep{1978ApJ...219L..85M}, is an eponym of the BL Lac category of blazars.  It is usually categorized as an LBL \citep{2018A&A...620A.185N}, but sometimes it is also classified as an IBL \citep{2011ApJ...743..171A}. During April 2021, the {\it Fermi}-LAT observed the historically largest $\gamma-$ray flare from BL Lac, which allowed us to search for the rapid $\gamma-$ray variations in the source. In this work, we report the first-ever detection of minute-timescale $\gamma-$ray variability in any BL Lac object by {\it Fermi}-LAT.

This paper is organized as follows. In Section \ref{sec:data} we give an overview of observations and the data analysis of {\it Fermi}-LAT. Results of our rapid $\gamma-$ray variability study are given in the Section \ref{sec:res}; Section \ref{sec:diss} presents the discussion and the summary of our work is given in the Section \ref{sec:summary}. 
\begin{figure}
\centering
\includegraphics[width=9cm, height=8cm]{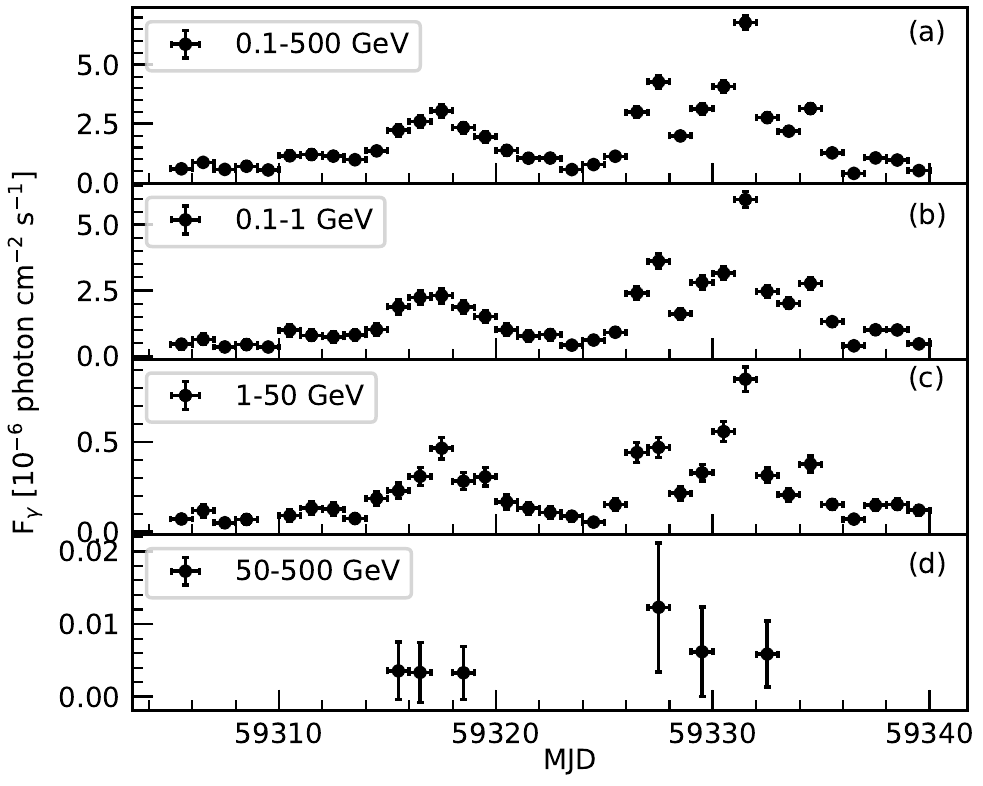}
\caption{\label{fig:lc_multi_bands} Daily-averaged $\gamma-$ray light curves of BL Lac in different energy bands. The energy range used to generate the light curve is mentioned in each panel.}
\end{figure}
\section{Observations and Data Analysis} \label{sec:data}
\begin{figure*}
\centering
\includegraphics[width=8.5cm, height=6cm]{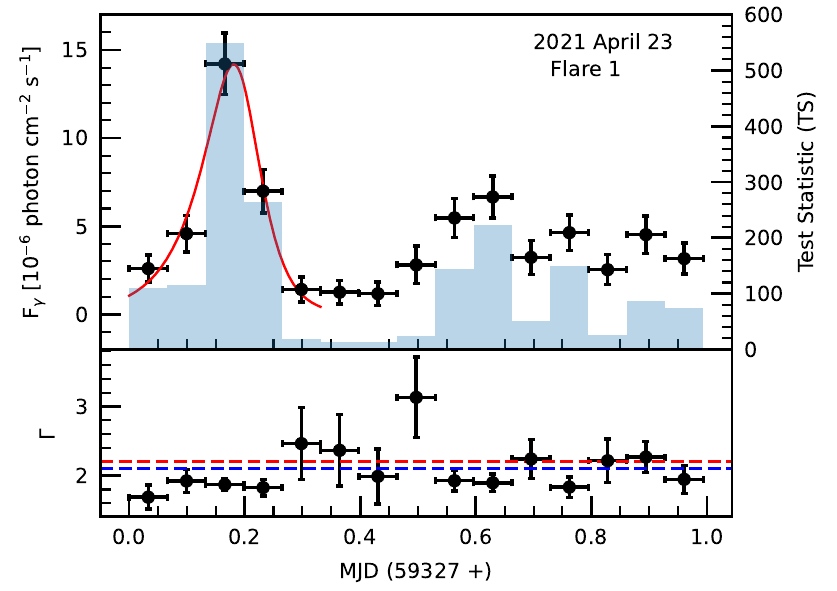} \includegraphics[width=8.5cm, height=6cm]{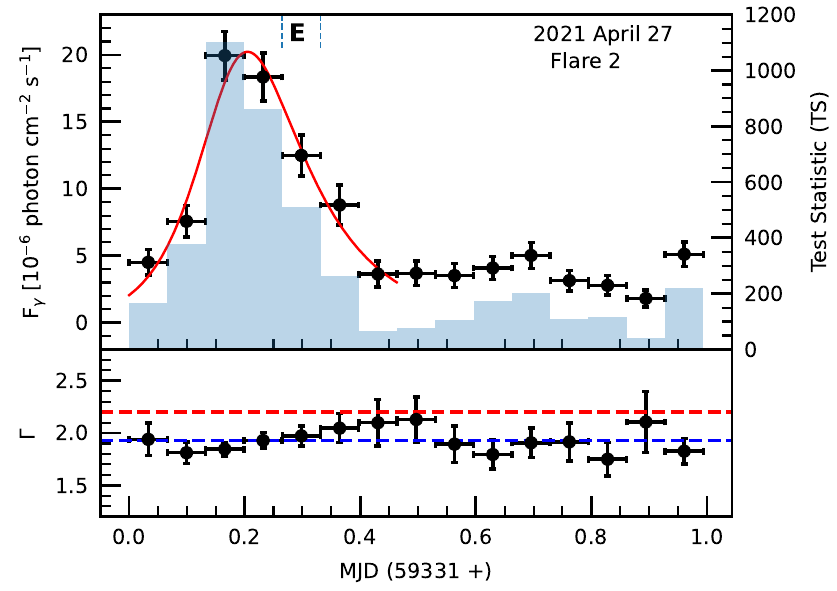}
\caption{\label{fig:lc_lat_A}The top panels are the orbit-binned $\gamma-$ray light curves of BL Lac for 2021 April 23 (left panel) and 2021 April 27 (right panel). The TS values for each time bin in both the light curves are shown as blue bars. The power-law index for each bin is plotted in the bottom panels. The red and blue horizontal lines in the bottom panels represent the 4FGL power-law index (2.2025) and the average photon index on that day, respectively.}
\end{figure*}
\subsection{Gamma-ray observations}
We used the Pass 8 (P8R3) {\it Fermi}-LAT $\gamma-$ray (0.1$-$500 GeV) data of BL Lac from MJD 59305 (2021 April 1) to MJD 59340 (2021 May 6). We analyzed the data following the standard LAT data analysis procedures\footnote{\url{https://fermi.gsfc.nasa.gov/ssc/data/analysis/}} and using the \Fermitool \ software package version 2.0.8 with the \IRF \ instrument response functions. For our analysis, we chose all the SOURCE class events (evclass=128 and evtype=3) within a circular region of 10 degrees (region of interest; ROI) around the blazar BL Lac. To select the good time intervals, we used a filter ``(DATA\_QUAL$>$0)\&\&(LAT\_CONFIG==1)'' and applied a maximum zenith-angle cut of 90 degrees to avoid the background $\gamma-$rays from the Earth's limb. We employed the unbinned maximum likelihood optimization technique for flux determination and spectral modelling \citep{2009ApJS..183...46A}. Our model file includes all the sources from the {\it Fermi}-LAT Fourth Source Catalog (4FGL; \cite {2020ApJS..247...33A}) within 20 degrees of the source as well as the Galactic and extragalactic isotropic diffuse emission components\footnote{\galactic \ and  \isotropic \ , respectively}. During the initial likelihood fit, all the parameters of the sources lying outside the ROI were kept fixed to their values in 4FGL, while the normalization and the spectral parameters of the sources within the ROI were left to vary freely. The normalizations of the diffuse emission components were also left free. 

\subsection{X-ray observations}
We generated the X-ray spectrum of BL Lac using the online {\it Swift}-XRT data products generator  tool\footnote{\url{https://www.swift.ac.uk/user_objects/}}(for details, see \cite{2009MNRAS.397.1177E}). This tool produces the pile-up corrected source spectrum, the background spectrum and the response files using the HEASOFT version 6.29. The X-ray spectrum of BL Lac was fitted with an absorbed power-law model in the XSPEC version 12.12.0 to obtain the 0.3-10 keV X-ray flux and the photon index. For the fitting, we assumed a fixed Galactic hydrogen column density of n$_H$ = 3.03 $\times$ 10$^{21}$ cm$^{-2}$ \citep{2013MNRAS.431..394W}. 

\begin{table*}
\centering
\caption{\label{tab:ffit}Results of the sum of exponential fit to the orbit-binned light curves. }
\begin{tabular}{cccccc}
\hline\hline
Flare    & Peak Time (MJD)   &  Peak Flux ($\times$ 10$^{-6}$ ph cm$^{-2}$ s$^{-1}$) & Rise time (T$_r$) & Decay time (T$_d$) & $\xi$\\\hline
Flare 1	 & 59327.19$\pm$0.02 & 13.61$\pm$1.84                                       & 1.42$\pm$0.32 hr    & 0.79$\pm$0.27 hr & -0.28\\
Flare 2       & 59331.18$\pm$0.02 & 19.47$\pm$1.67   				    & 1.49$\pm$0.29 hr    & 2.63$\pm$0.47 hr  & 0.28 \\\hline
\end{tabular}
\end{table*}

\begin{figure*}
\centering
\includegraphics[width=8.5cm, height=6cm]{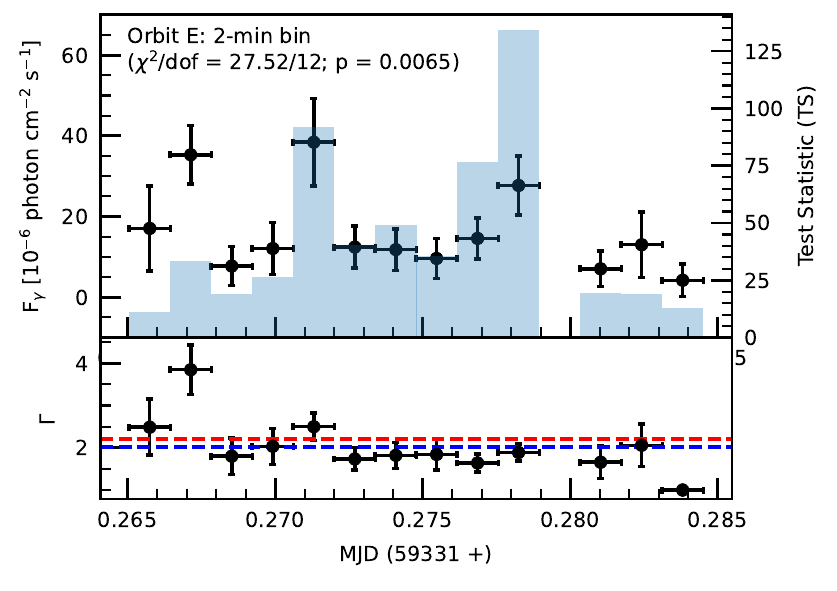} \includegraphics[width=8.5cm, height=6cm]{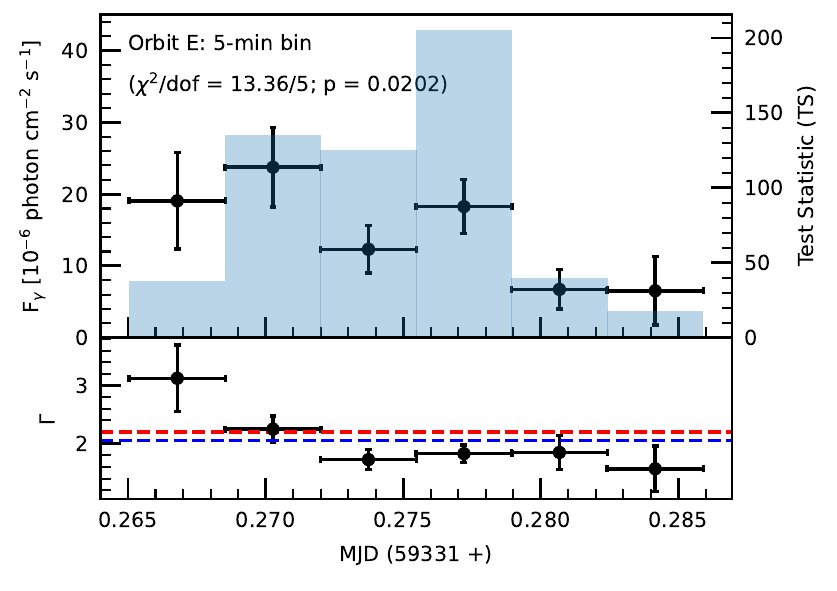}
\caption{\label{fig:min_scale_var}The top panels are the 2-min (left) and 5-min (right) binned light curves for the orbit E. The TS values for each time bin in both the light curves are shown as blue bars. The power-law index for each bin is plotted in the bottom panels. The red and blue horizontal lines in the bottom panels represent the 4FGL power-law index (2.2025) and the average photon index on that day, respectively.}
\end{figure*}

\section{Results}\label{sec:res}
\subsection{$\gamma-$ray light curve}\label{subsec:lc}
Figure \ref{fig:lc_lat} presents the $\gamma-$ray light curves of BL Lac for a period of about one month that includes the brightest outburst observed on 2021 April 27. Although the spectral shape of BL Lac is defined as log-parabola (LP) in the 4FGL catalog \citep{2020ApJS..247...33A}, we generated the light curves by modeling the spectra in each time bin as a simple power-law (PL), since the PL indices have smaller statistical uncertainties than those obtained from the complex LP model \citep[e.g.][]{Abdo_2011}. Also, choosing a simple PL model is more appropriate for this work as we are probing the shortest timescale $\gamma-$ray variations, thereby dealing with lesser photon statistics. The arrival time and the energy of the highest energy photon (bottom panel) coming from the source were derived using the tool \srcprob \ on ULTRACLEAN event class (evclass=512).

The maximum 1-day averaged flux (above 100 MeV) was observed on MJD 59331 (2021 April 27) reaching (6.8$\pm$0.3) $\times$ 10$^{-6}$ photons cm$^{-2}$ s$^{-1}$, which is the highest daily binned flux ever observed from this source. The photon index corresponding to this highest flux is 1.90$\pm$0.03, which is slightly harder than the value (2.2025) mentioned in the 4FGL catalog. The 3-hr binned $\gamma-$ray light curve of BL Lac, shown in panel (c) of Figure \ref{fig:lc_lat}, indicates two sharp $\gamma-$ray flares.

To investigate any instrumental uncertainties in the analysis, we also generated the daily-binned light curves of BL Lac in 0.1-1 GeV, 1-50 GeV, and 50-500 GeV energy bands for the period considered. As seen in Figure \ref{fig:lc_multi_bands}, the light curves in those three energy bands follow a pattern similar to the total 0.1-500 GeV energy band light curve. There are only six data points in the 50-500 GeV light curve as the source was not significantly (TS >9) detected in this energy range in other time bins.

The highest-energy photon, 128 GeV, was recorded with $>$99.99\% probability on MJD 59333 (2021 April 29), which is at the decline phase of the main flare. A similar trend was also seen in 3C 279 where the highest energy photon was observed at the end of the outburst phase \citep{2016ApJ...824L..20A}.

\subsection{Sub-orbital timescale variability}\label{subsec:var}
As shown in panel (c) of Figure \ref{fig:lc_lat}, the source flux exceeded the value of 10$^{-5}$ photons cm$^{-2}$ s$^{-1}$ on two days namely, 2021 April 23 (MJD 59327) and 2021 April 27 (MJD 59331) on three occasions with high photon statistics; MJD 59327.18756 (TS = 798.985), MJD 59331.18756 (TS = 1968.951), and MJD 59331.31256 (TS = 776.058). This allowed us to further resolve the light curve with shorter timescales on these two days. We first generated the light curves with bin size equal to the orbital period ($\sim$95.4 minutes) of the {\it Fermi}-LAT. The orbit-binned light curves of BL Lac on 2021 April 23 (left) and April 27 (right) are shown in the top panels of Figure \ref{fig:lc_lat_A}.  We estimated the shortest flux doubling/halving timescales on these two epochs as follows:
\begin{equation}\label{eq:doubling_time}
    F(t_{2}) = F(t_{1})\times2^{\Delta t/\tau}
\end{equation}
Here, F($t_{1}$) and F($t_{2}$) are the flux values at times $t_{1}$ and $t_{2}$ respectively, $\Delta$ t = $t_{2}$-$t_{1}$ and $\tau$ denotes the flux doubling/halving timescale. We observed a flux halving timescale of $\sim$(0.68$\pm$0.22) hr with a significance of $\sim$4$\sigma$ during the decay of the flare on MJD 59327. We also detected a flux doubling timescale of $\sim$(1.14$\pm$0.21) hr with $\sim$6$\sigma$ significance during the rise of the flare on MJD 59331.

To understand the temporal evolution of the flux, we fitted the peaks of orbit-binned light curves by a function of the sum of exponential defined as \citep{2010ApJ...722..520A}
\begin{equation}
    F(t) = 2 F_0 \left(e^{\frac{t_0-t}{T_r}} + e^{\frac{t-t_0}{T_d}}  \right)^{-1} 
\end{equation}
Here, $F_0$ is the flux value at time $t_0$ denoting the flare amplitude, $T_r$ is the rise time, and $T_d$ is the decay time of the flare.
The results of the fit are given in Table \ref{tab:ffit}. We also estimated a parameter $\xi = (T_d - T_r)/(T_d + T_r)$ that describes the symmetry of the flares \citep{2010ApJ...722..520A}. For both the flares,  we found  $-0.3 < \xi < 0.3$ implying that these flares are symmetric. 

Rapid flux variations with high photon statistics observed on MJD 59327 and MJD 59331 provide us with an opportunity to examine ultra-fast flux variations on the timescale of a few minutes. To detect such rapid flux variations, we generated 2-min, 3-min and 5-min binned light curves for each orbit on these two days. Similar to \cite{2016ApJ...824L..20A} and \cite{2018ApJ...854L..26S}, we searched for minute-scale variability on these two days by fitting a constant flux to each orbit for all three time bins and subsequently, computing the probability ($p-$value) of the flux to be constant. Since the detection of minute-timescale $\gamma-$ray variations are very rare, we conservatively chose 95\% confidence level (p-value smaller than 0.05) as the detection limit for sub-orbital variability.
On MJD 59327, the $p$-values for all the orbits and for all the time bins are found to be consistent with constant flux. Similarly, on MJD 59331, for all the orbits, except orbit E, we obtained $p-$values denoting no flux variations for all the time bins. 
For orbit E, we detected minute-scale variability in the 2-min binned ($p$=0.0065, $\chi^2$/dof = 27.52/12) and 5-min binned ($p$=0.0202, $\chi^2$/dof = 13.36/5) light curves. However, in the 3-min binned light curve of orbit E the variations were not significant ($p$=0.6339, $\chi^2$/dof = 6.12/8). Our $p-$values are comparable to those found by \cite{2016ApJ...824L..20A} in a single orbit, during a giant flare of blazar 3C 279. The 2-min binned and 5-min binned light curves for orbit E are shown in the top panels in Figure \ref{fig:min_scale_var}. We also searched for the flux doubling/halving timescales in these light curves using Equation \ref{eq:doubling_time}. We found a halving time scale of $\sim$ (1$\pm$0.3) minute with $\sim$3.2$\sigma$ in the 2-min binned light curve of orbit E. 
\subsection{Gamma-ray spectral variability}\label{subsec:spec_var}
\begin{figure}
\centering
\includegraphics[width=8cm, height=7cm]{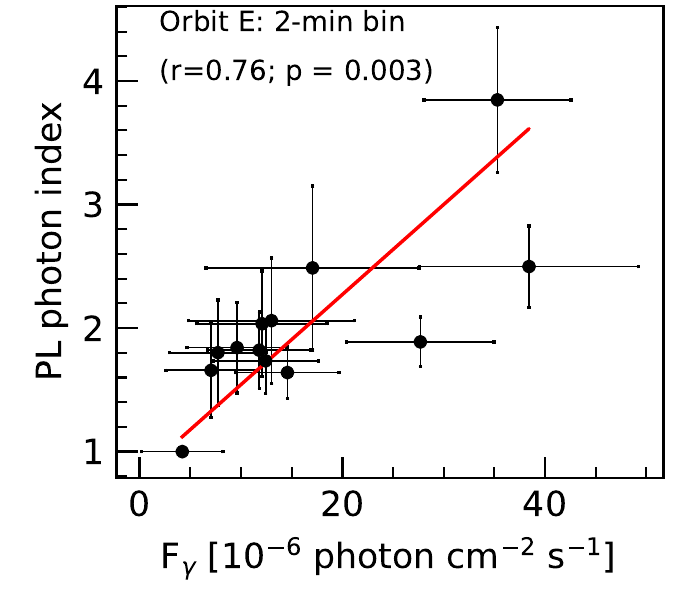} 
\caption{\label{fig:flx_indx_2min}Variation of PL photon index with $\gamma-$ray flux for the 2-min binned light curve of orbit E. A softer-when-brighter trend is clearly visible. }
\end{figure}

We investigated the $\gamma-$ray spectral variability of BL Lac on different time bins, namely daily, orbital and shortest time bins. For all the time bins, the average PL photon indices are somewhat harder than its value (2.2025) in the 4FGL catalog. The average PL photon indices for these time bins are given in Table \ref{tab:indx_var}. We also searched for any correlation between the observed $\gamma-$ray flux and the PL index on these time bins using the Pearson correlation. The results of the correlation study are given in Table \ref{tab:indx_var}. As seen from Table \ref{tab:indx_var}, no significant correlation was found between the $\gamma-$ray flux and PL index on any of the time bins except on the 2-min time bin. We found a significant ($p<$0.01) positive correlation between $\gamma-$ray flux and PL index in the 2-min binned light curve indicating a softer-when-brighter trend. The variation of PL photon index with $\gamma-$ray flux for 2-min binned light curve of orbit E is shown in Figure \ref{fig:flx_indx_2min}.

\begin{table}
\centering
\caption{\label{tab:indx_var}Results of spectral variability of BL Lac on different time bins. Here, $r$ and $p$ represent the Pearson correlation coefficient and the null hypothesis probability, respectively.}
\begin{tabular}{lccc}
\hline\hline
Bin size & Average PL index   & \multicolumn{2}{c}{Flux vs PL index}   \\\hline
         &                      &     r    &      p             \\\hline
1-day         &  1.95$\pm$0.02                  &      -0.16 & 0.349           \\
1-Orbit (flare 1)   &  2.11$\pm$0.08	&	-0.38 & 0.161		\\     
1-Orbit (flare 2)    &  1.93$\pm$0.04	&	-0.14 & 0.622		\\
2-min (orbit E)            &  2.02$\pm$0.11	&	0.76 & 0.003		\\
5-min (orbit E)            &  2.06$\pm$0.14	&	0.59 & 0.209		\\
\hline
\end{tabular}
\end{table}

\section{Discussion} \label{sec:diss}
For the first time, a rapid (minute-timescale) $\gamma-$ray variability is detected in any BL Lac category of blazars by {\it Fermi}-LAT. We observed significant variations in the minute-scale binned $\gamma-$ray light curves with a halving timescale of $\sim$1 minute during the historically bright $\gamma-$ray flare from BL Lac on MJD 59331. The detection of such a rapid variability timescale challenges the existing $\gamma-$ray emission models.

The black hole mass for BL Lacertae is 1.7 $\times$ 10$^8$ M$_{\odot}$ \citep{2014Natur.510..126Z} and the corresponding event horizon light-crossing time is $\sim$14 minutes. The detected variability timescale ($\sim$ 1 minute) is much shorter than the event horizon light-crossing time indicating that the enhanced $\gamma-$ray emission is coming from a very compact region within the jet. Such rapid variations could be triggered either by dissipation in a small fraction of the black hole magnetosphere at the base of the jet or by small scale instabilities within the jet \citep{2008MNRAS.384L..19B}. Alternatively, the jet could be much more extended and the emission can come from a localized region in the
jet much smaller than the width of the jet itself. In this scenario too, the innermost region of the jet at the sub-parsec level is responsible for the observed GeV emission.

The total jet power that is required to produce the observed $\gamma-$ray luminosity $L_{\gamma} \sim$ 5 $\times$ 10$^{47}$ erg s$^{-1}$ is  $L_j \simeq L_{\gamma}/(\eta_j\Gamma^2)$, where $\eta_j \sim 0.1$, is the radiative jet efficiency \citep{2016ApJ...824L..20A}. The jet power should be less than the Eddington luminosity $L_{Edd} \sim$ 2 $\times$ 10$^{46}$ erg s$^{-1}$ which implies that $\Gamma > 16$.

The minimum Doppler factor, $\delta_{min}$, of the $\gamma-$ray emission region can also be estimated numerically using a $\delta-$function approximation for the $\gamma\gamma$ opacity constraint and the detected high energy $\gamma-$ray photons \citep{10.1093/mnras/273.3.583,2010ApJ...716.1178A}. Assuming that the $\gamma-$rays are produced via the SSC scattering process and the target photons for SSC are X-ray photons, a lower limit of the Doppler factor can be calculated as

\begin{equation}
    \delta_{min} = \left[\frac{\sigma_T d^2_L (1+z)^2 f_{\epsilon} E_1}{4 \tau m_e c^4}\right]^{1/6}
\end{equation}
where $\sigma_T$ is the Thomson scattering cross section, d$_L$ = 307 Mpc is the luminosity distance\footnote{Assuming $H_0$=71 km s$^{-1}$ Mpc, $\Omega_M$=0.27, $\Omega_{\Lambda}$=0.73 \citep{2011ApJS..192...16L}} for BL Lac, $f_{\epsilon}$ = 7 $\times$ 10$^{-11}$ erg cm$^{-2}$ s$^{-1}$ is the X-ray flux obtained from the {\it Swift}-XRT observation (obsid 00034748061; X-ray photon index = 2.28$\pm$0.08), and $E_1$ = 26 GeV/$m_ec^2$ is the highest energy photon. We estimated $\delta_{min}$ to be $\sim$ 15, which is consistent with the $\delta_{min}$ = 13-17 obtained by \cite{2013ApJ...762...92A}. 

Taking $\delta$=16, the detected size of the $\gamma$-ray emitting region $R \leq c \tau \left(\delta/(1+z)\right)$ $\leq  2.69 \times 10^{13}$ cm. For a conical geometry of the jet, the distance of the $\gamma-$ray emission region from the central super-massive black hole is $D \leq (2 c \Gamma^2 \tau)/(1+z)$ $\approx$ 8.62 $\times$ 10$^{14}$ cm, assuming $\Gamma = \delta_{min}$ \citep{Abdo_2011}.

Flares seen on the orbit-binned light curves have a symmetric profile and can thus be associated with the crossing time of radiation through the emitting region or can be explained by the superposition of several short-duration flares \citep{2010ApJ...722..520A}. 

\section{Summary} \label{sec:summary}
In this work, we report the first detection of minute-timescale GeV $\gamma-$ray variability in the BL Lac category of blazars by {\it Fermi}-LAT. We detected a flux halving timescale of $\sim$ 1 minute from BL Lac on 2021 April 27. This observed short timescale of variability requires a minimum bulk Lorentz factor of 16 to have the jet power lesser than the Eddington value. Also, $\gamma\gamma$ transparency argument for that epoch of short timescale variability detection requires a minimum Doppler factor of $\sim$ 15. We found a softer-when-brighter trend in the 2-min binned light curve of orbit E.

\begin{acknowledgements}
We thank the referee for his/her critical comments that helped in the improvement of the manuscript. We thank Dr Vaidehi S. Paliya for a fruitful scientific discussion. This work has made use of Fermi data collected from the Fermi Science Support Center (FSSC) supported by the NASA Goddard Space Flight Center. This research has made use of the High-Performance Computing (HPC) resources (\url{https://www.iiap.res.in/?q=facilities/computing/nova}) made available by the Computer Center of the Indian Institute of Astrophysics, Bangalore. This work made use of data supplied by the UK Swift Science Data Centre at the University of Leicester.

\end{acknowledgements}
\bibliographystyle{aa} 
\bibliography{ref} 

\end{document}